\documentclass[fleqn,twoside]{article}
\usepackage{espcrc2}

\usepackage{graphicx}
\usepackage[figuresright]{rotating}

\newcommand{\lsim}   {\mathrel{\mathop{\kern 0pt \rlap
  {\raise.2ex\hbox{$<$}}}
  \lower.9ex\hbox{\kern-.190em $\sim$}}}
\newcommand{\gsim}   {\mathrel{\mathop{\kern 0pt \rlap
  {\raise.2ex\hbox{$>$}}}
  \lower.9ex\hbox{\kern-.190em $\sim$}}}
\def\be{\begin{equation}}
\def\ee{\end{equation}}    
\def\ba{\begin{eqnarray}}
\def\ea{\end{eqnarray}}

\def\nuebar{\bar{\nu}_e}

\def\nue{\nu_e}
\def\nux{\nu_x}
\def\numu{\nu_\mu}
\def\nutau{\nu_\tau}
\def\Dm2{\Delta m^2}

\newcommand{\AmS}{{\protect\the\textfont2
  A\kern-.1667em\lower.5ex\hbox{M}\kern-.125emS}}

\title{Solar and Reactor Neutrinos:\\
Upcoming Experiments and Future Projects}

\author{Stefan Sch\"onert \address{Max-Planck-Institut f\"ur Kernphysik,\\
        Saupfercheckweg 1, 69117 Heidelberg, Germany}}

\begin{document}

\begin{abstract}
Sub-MeV solar neutrino experiments and long-baseline reactor
oscillation experiments toe the cutting edge of neutrino research.
The upcoming experiments KamLAND and BOREXINO, 
currently in their startup  and final construction phase respectively, 
will provide essential information on neutrino properties
as well as on solar physics. Future projects, at present under
development, will measure the primary solar neutrino fluxes via 
electron scattering and neutrino capture in real time. 
High precision data for lepton mixing 
as well as for stellar evolution theory will become available in the 
future. This paper aims to give an overview of the 
upcoming experiments and of the projects under development.

\vspace{1pc}
\end{abstract}

\maketitle

\section{Introduction}

In the past, experiments with solar neutrinos and experiments with 
neutrinos from nuclear reactors probed different ranges
of the parameter space for neutrino oscillations.
Basically, the  distances between source and detector in reactor neutrino 
experiments ($\lsim 1$~km) were too short to probe the 
parameter space relevant to explain the solar neutrino deficit 
\cite{till,concha}. In the very near future, with the upcoming experiments 
KamLAND \cite{kamland} and BOREXINO
\cite{borexino}, it will be possible to study the parameter space
of the large mixing angle (LMA) solution not only with solar, 
but as well with reactor neutrinos.
Although BOREXINO is optimized for solar neutrino detection, 
it will be able to detect
$\nuebar$'s from far distant European nuclear reactors 
\cite{stefanTaup97}. {\em Vice versa}, 
KamLAND is dedicated to reactor neutrino spectroscopy,
however, if radio impurities are sufficiently low, 
it will also measure solar neutrinos. 

Future projects, for example LENS \cite{lens} and
XMASS \cite{xmass}, seek to measure 
the primary pp-, $^7$Be- and CNO neutrino fluxes 
in real time via neutrino capture and neutrino-electron scattering 
respectively. Such measurements will provide
important information on the flavor of solar neutrinos,
on mixing parameters and on solar physics. 
If the LMA solution
is realized in nature, $\Dm2$ will be determined with high
accuracy using reactor neutrinos. However, the measurement of the 
mixing angle will be statistically limited.    
Precision measurements of the pp- and $^7$Be fluxes by 
elastic scattering (ES) or neutrino capture (CC) would 
improve considerably the accuracy of the determination of the 
mixing angle. If the ``LOW'', ``(quasi-) VAC'' or 
small mixing angle (SMA) is the correct solution 
(see Fig.~1 for terminology), then only solar 
neutrino experiments can explore the parameter spaces. 
Full information of the solar neutrino spectrum,  via both 
ES and neutrino capture will be essential in any of the
mentioned scenarios.

\section{Objectives of future experiments}

Since flavor conversion has been established from the 
the combined analysis of SNO and Super-Kamiokande (SK) data \cite{McDonald}, 
the scientific goals of upcoming experiments and future projects
shift from discovery and identification of the deficit to
a comprehensive study of the phenomena: 
first, to fix the specific region of the 
parameter space, and second, to measure with high accuracy the 
neutrino oscillation parameters and the primary solar neutrino fluxes.

\subsection{Test of stellar evolution theory} 

Solar neutrinos are a unique tool to directly study the fusion
processes in the interior of the sun.
In particular, {\em direct} measurements, with energy
information and in real time, of neutrinos from the primary 
pp-, $^7$Be- and CNO processes are of fundamental interest.   
They will provide complementary information to observational data from 
helioseismology and probe the predictions of solar models.

\subsection{Neutrino properties}

One of the main goals of future experiments
is to study neutrino oscillations and determine
the oscillation parameters. In particular they need to:
\begin{itemize}
\item confirm $\nue$ conversion seen by SK and SNO at sub-MeV
  energies;
\item observe directly the signature of oscillations, i.e. time
  variations or energy spectrum deformations;
\item probe flavor composition at sub-MeV energies (active
  vs. sterile neutrinos, sterile contributions);
\item establish the correct oscillation solution 
(LMA, LOW, (quasi-)VAC, SMA) and measure the mixing parameters
with high accuracy.
\end{itemize}
\noindent 

Other questions comprise: the investigation of 
$\Theta_{13}$,  magnetic moment searches, coherent
scattering off nuclei and neutrino instability.

\begin{figure}
\label{survival}
\includegraphics[width=15pc]{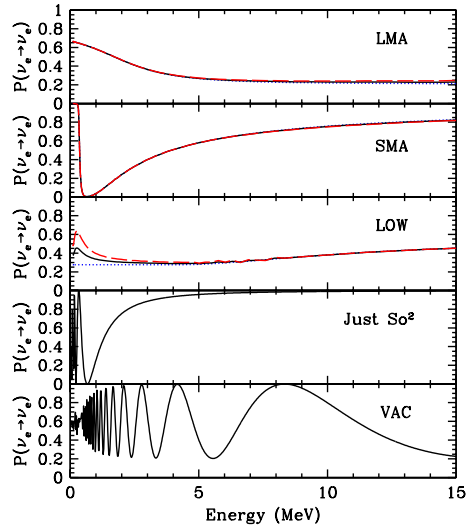}
\vspace{-1.5truecm}
\caption{Energy dependent survival probabilities for 
solar neutrinos for various oscillation scenarios  
\cite{BahKrastSmir}.
The characteristic difference of the 
various solutions at sub-MeV energies should be noted.}
\end{figure}

\section{Sub-MeV signature of solar $\nu$-oscillations}
At $^8$B $\nu$-energies 
the various oscillation solutions differ only slightly with 
respect to the energy dependence of the survival probability. 
At sub-MeV energies, however, they differ dramatically. 
Moreover, for the LOW and VAC solutions the survival probability
varies periodically in time for sub-MeV neutrinos. 
Day/night variations will be observable in the case of the LOW solution
and variations on the scale of weeks to months for VAC solutions. 
Fig.~1 displays the characteristic 
survival curves for various oscillation scenarios.
The ``Just So$^2$'' and the ``SMA'' solutions are disfavored when
including the recent results of SNO.

\section{Experiments and R\&D projects}

Table~\ref{tab:projects} lists sub-MeV solar  as 
well as reactor neutrino experiments which soon become operational, 
as well as projects which are currently in an early stage of  
development.
 
\begin{table*}[htb]
\caption{Listing of upcoming experiments and projects under development.}
\label{tab:projects}
\newcommand{\m}{\hphantom{$-$}}
\newcommand{\cc}[1]{\multicolumn{1}{c}{#1}}
\begin{tabular}{lllll}
\hline
Project   & Goals   & Method & Technique (Target)  & Status \\
\hline
KamLAND \cite{kamland} & 1) $\nuebar^{\rm reactor}$ & CC &
                                                       LS (CH) & 2001/2 \\
                       & 2) $\nu^{sun}:^7$Be, $^8$B &
                         ES & LS (CH) &  \\
BOREXINO \cite{borexino}& 1)$\nu^{sun}$: $^7$Be, pep, $^8$B 
                                 &  ES & LS (CH) & 2002/3 \\
                        & 2)  $\nuebar^{\rm reactor}$ & CC &
                                                       LS (CH) &2002/3\\
LENS \cite{lens}    & $\nue^{sun}$: pp, $^7$Be, CNO 
                    & CC & LS (CH+metal) & R\&D: 2002/3 \\
XMASS \cite{xmass}  & $\nux^{sun}$: pp, $^7$Be
                    & ES  & LS (Xe) & R\&D: 2002/3 \\
HERON \cite{heron}  & $\nux^{sun}$: pp  & ES  & LS (He)   & R\&D \\ 
CLEAN \cite{clean}  & $\nux^{sun}$: pp  & ES  & LS (He,Ne)& R\&D \\ 
MOON \cite{moon}    & $\nue^{sun}$: pp, $^7$Be  
                    & CC & hybrid or LS (CH+metal) & R\&D\\
TPC \cite{tpc,munu} &  $\nux^{sun}$: pp, $^7$Be 
                    & ES  & TPC (He+CH) & R\&D \\
GENIUS \cite{genius}&  $\nux^{sun}$: pp 
                    & ES  & Germanium & R\&D \\
LOW-C14 \cite{lowc14}&  $\nux^{sun}$: pp 
                    & ES  & LS (CH) & R\&D \\
\hline
Kr2Det\cite{kr2det} & $\nuebar^{\rm reactor}$: U$_{e3}$ & CC &
                                                       LS (CH) & R\&D \\
TEXONO\cite{texono} &$\nuebar^{\rm reactor}$: $\mu_\nu$ & ES 
                    & solid scint. & R\&D \\ 
\hline
\end{tabular}\\[2pt]
ES: Elastic scattering of neutrinos off electrons (CC+NC); CC:
neutrino capture (charged current; CC); LS: liquid scintillator; CH: hydro-carbon.
\end{table*}

\subsection{ES and CC-detection of sub-MeV neutrinos}

\begin{figure}
\label{fig:BX-Lens}
\includegraphics[width=20pc]{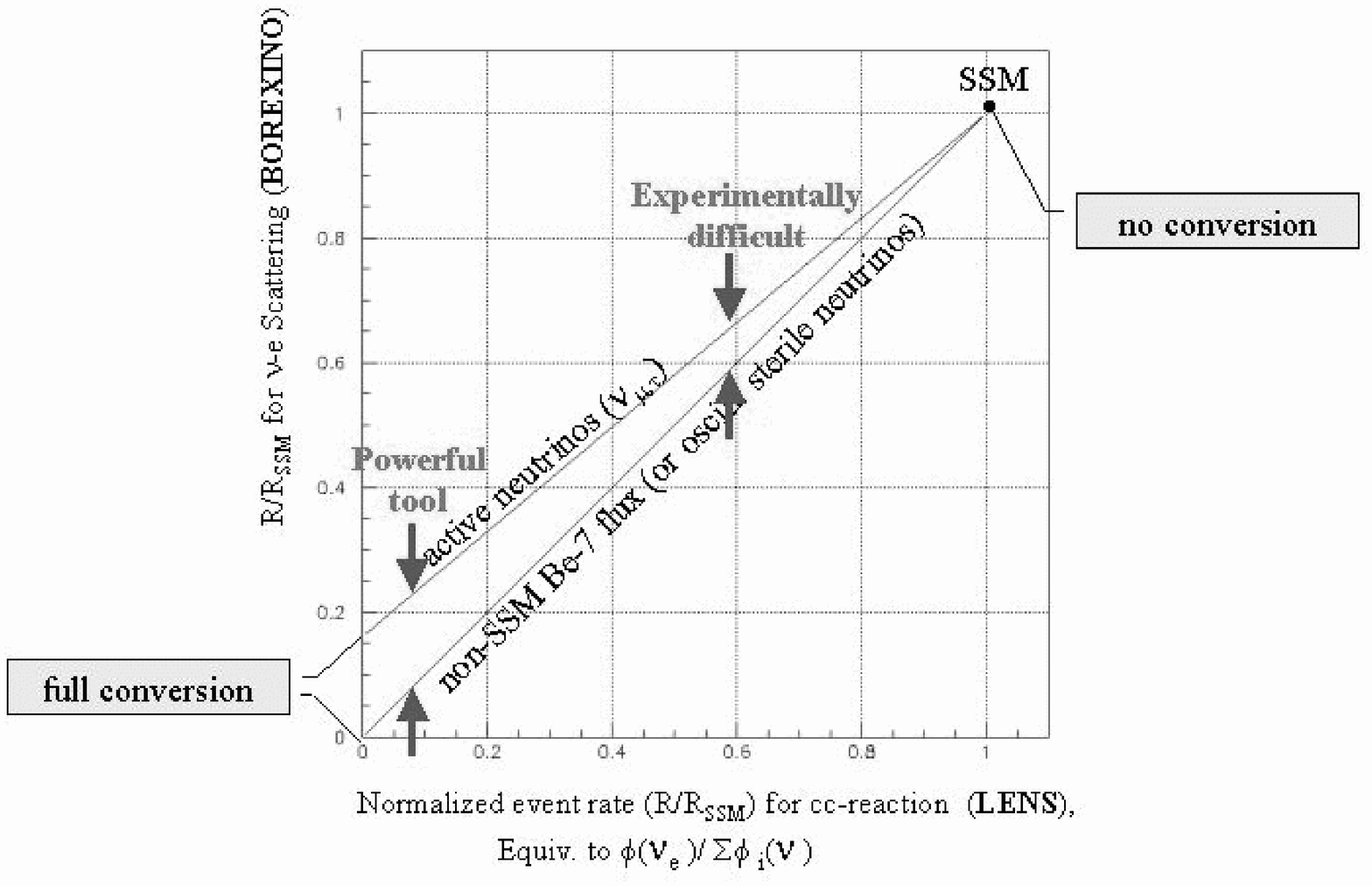}
\vspace{-3.truecm}
\caption{}
\end{figure}

Two different types of interactions will be exploited
for the detection of sub-MeV real-time measurements: 
$\nux e$--scattering (ES) which involves charged (CC) and neutral (NC)
current interactions, and neutrino capture which only
occurs via CC-interaction. The first is sensitive 
to all active flavors -- albeit $\numu, \nutau$ interact 
with a probability of about 1/6 of $\nue$ -- while the latter is sensitive 
purely to $\nue$. Combining the results of an ES experiment with
those of a neutrino capture experiment provides  
a unique neutral current probe.
Currently this provides the only possibility to study the contributions
of $\nu_{\mu}$ and $\nu_{\tau}$ in the primary pp- and $^7$Be neutrino
fluxes. Electron scattering
and neutrino capture experiments are therefore complementary,  
and should  be realized 
in a coordinated strategy in order to maximize the potential physics 
output. For purposes of illustration only, Fig.~2 shows 
the expected relative interaction rates of $^7$Be neutrinos in
BOREXINO and LENS. It should be noted that modest suppression factors, 
as for example expected for the the LMA solution, 
challenge the accuracy reachable in the 
experiments. For instance, in the particular case indicated in 
Fig.~\ref{fig:BX-Lens} with ``experimentally difficult'', an error 
of $\lsim3\%$ is needed both for BOREXINO
and LENS in order to obtain a signal for 
$\numu,\nutau$ appearance $\gsim 3\sigma$~CL.

\subsection{BOREXINO and KamLAND}
As illustrated in Figs.~3 and 4, 
the BOREXINO (BX) and
KamLAND (KL) detectors are similar in design. BX is optimized for solar 
neutrino detection, thus design priority has been given to 
high radiopurity, while KL is optimized for reactor neutrino
detection where radiopurities are less stringent.  
Both detectors use liquid scintillators as $\nu$-target and 
detection medium. Reactor-$\nuebar$'s are detected via
$\nuebar p \to e^+ n$ where $e^+$ carries the energy information
of the interacting neutrino. 
The $\nuebar$-tag is given by a coincident $e^+ - {\rm n}$ signal
which is correlated in space and separated in time with 
$\tau=200 \mu$sec.
Solar neutrinos are measured via $\nux e^-$-scattering
($x=e,\mu,\tau)$. 
The Compton-like recoil electron provides the energy information
of the interaction. 
About 55 events/day/100 tons are expected in the ``$^7$Be neutrino 
window'' (i.e. recoil electrons
with energies between 250-800 keV). 
Since  $\nux e^-$--signals cannot
be distinguished from radioactive background as $\beta$-decay
events with the same energy, ultra-low backgrounds are required
for neutrino detection. The signal and background spectra
expected in BOREXINO are displayed in Fig.~5. 
Expected detector performance and specifications for  
KL and BX  are summarized in Tab.~\ref{tab:BXKL}. 
KL needs to meet similar radiopurity specifications to achieve 
solar-$\nu$ detection. However, variable time effects, such 
as day/night variations,
can be observable even with a background larger than $10^{-16}$g/g. 
KL data taking commenced the end of 2001 and the first results from 
reactor $\nuebar$-measurements are expected in 2002. BX 
scintillator filling will be completed early  2003 
and first solar neutrino results might be available during the same
year.  

\begin{table}[htb]
\caption{Features of BOREXINO and KamLAND}
\label{tab:BXKL}
\newcommand{\m}{\hphantom{$-$}}
\newcommand{\cc}[1]{\multicolumn{1}{c}{#1}}
\begin{tabular}{lll}
\hline
   & BX   & KL \\
\hline
Mass:    & 300 t     & 1 kt \\
LS:              & PC+PPO    & MO+PC+PPO \\
Yield [pe/MeV]: & ~430  & ~300 \\
Purities:     &       &      \\
~U,Th [g/g]:    &$\le {10^{-16}} ^\ast $  & $\le
{10^{-14}}^{\ast\ast}$ \\
~K [g/g]:       & $\le {10^{-14}}^\ast$  & - \\
~$^{14}$C/$^{12}$C  & $\le 10^{-18}$ & -  \\
CP:   & 1$\times$GS  & 7$\times$GS \\
Reactor dist.:  & ~800 km       & ~160 km \\   
\hline
\end{tabular}\\[2pt]
LS: Liquid scintillator; PC: pseudocumene; MO: mineral oil; 
pe: photo electrons; $^\ast$: for solar-$\nu$ detection,
 $^{\ast\ast}$: for reactor-$\nuebar$'s, CP: cosmogenic production
at Kamioka is a factor 7 higher than at Gran Sasso.  
\end{table}

\begin{figure}
\label{fig:BX-det}
\includegraphics[width=18pc]{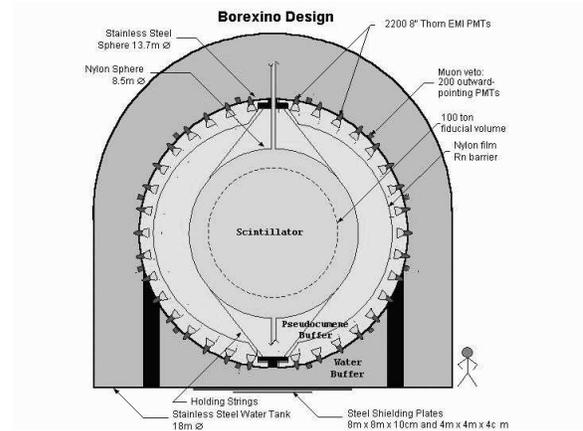}
\vspace{-1.0truecm}
\caption{Schematic view of the BOREXINO detector at Gran Sasso, Italy.}
\end{figure}

\begin{figure}
\label{fig:KL-det}
\includegraphics[width=13pc]{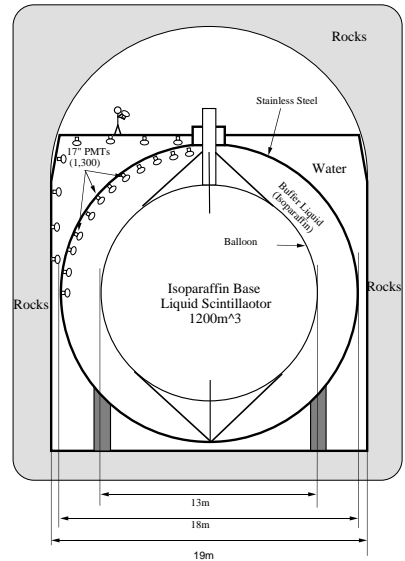}
\vspace{-1.0truecm}
\caption{View of the KamLAND detector at Kamioka, Japan.}
\end{figure}

\begin{figure}
\label{fig:BX-spec}
\includegraphics[width=22pc]{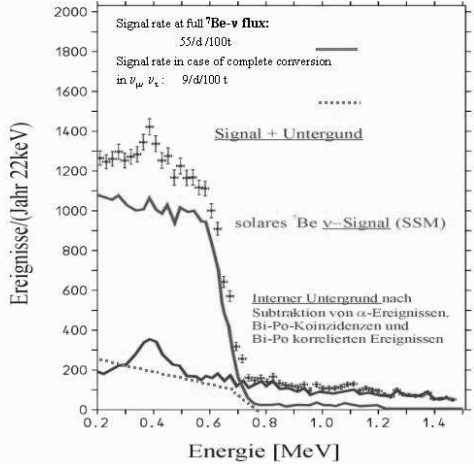}
\vspace{-2.0truecm}
\caption{Electron recoil spectrum of $^7$Be-$\nu$'s and 
expected background in BOREXINO.}
\end{figure}

The reactor signal in KL (and as well in BX) will be strongly
suppressed in the case where  LMA parameters drive solar neutrino oscillations.
The oscillation length 
$L[{\rm m}]=1.24 \cdot E [{\rm MeV}] /\Dm2 [{\rm eV^2]}$
would amount to 100 km  
for 3~MeV 
reactor--$\nuebar$'s and 
$\Dm2=3.7\times 10^{-5}~{\rm eV}^2$. 
Fig.~6 shows the parameter space in
which an observable rate suppression is expected. In case of 
no oscillations, KL will measure about 800 events per year
in comparison to 30 events in BX. This obviously translates
to a higher sensitivity with respect to $\sin^2 2\Theta$.
The recent best fit value $\sin^2 2\Theta =0.79$, 
however, should provide unambiguous
signals in both experiments provided that background levels
are as expected. 

For $\Dm2$ values between $\sim 1\times 10^{-5}{\rm eV}^2$ and 
$\sim 2\times 10^{-4}{\rm eV}^2$, KL will not only observe 
a suppression of the rate, but also deformation of the
shape of the  well known positron energy spectrum. From the shape
one can then infer, with high accuracy, the actual value
of $\Dm2$. For values larger $2\times 10^{-4}{\rm eV}^2$
the oscillation signal smears out and only a lower limit
can be given by KL. The upper ``hard core'' limit then will be given by 
the CHOOZ experiment at around $1\times 10^{-3}{\rm eV}^2$.
A follow-up experiment would be needed in this particular
case. A a reactor experiment with a baseline of about 20 km
is under discussion  to address this question \cite{Em4}.

\begin{figure}
\label{fig:KL-BXexcl}
\includegraphics[width=15pc]{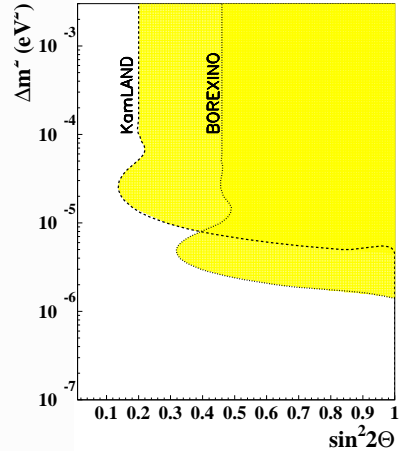}
\vspace{-1.0truecm}
\caption{Sensitivity of KL  and BX (2$\sigma$ CL) 
for $\nu$-oscillation search with reactor-$\nuebar$. 
The shaded area would provide a measurable suppression 
of the $\nuebar$-rate based on 3 years of data taking. 
KL's sensitivity  increases to $\sin^2 2\Theta = 0.1$ 
in the case of  background free detection of $\nuebar$'s.}
\end{figure}


\subsection{ES experiments for sub-MeV $\nu$ detection}

Various $\nu e^-$--scattering experiments for pp neutrino
detection are under development. With an interaction rate 
of about 2/ton/day for pp-$\nue$, the detectors need to have 
a target mass of about 10 tons to acquire high statistical accuracy.
 $^{14}$C/$^{12}$C ratios at $10^{-18}$ in organic scintillators, 
as determined with the CTF of BX \cite{BXC14}, 
prohibit the measurement of pp-neutrinos
since the 156~keV $\beta$-decay endpoint significantly
obscures the pp-energy range. If a low-$^{14}$C liquid 
scintillator  with $^{14}$C/$^{12}$C $\lsim10^{-20}$ 
could be produced, as discussed in \cite{lowc14},
a BX type detector could be realized for pp-$\nu$ detection.
A promising approach to overcome the $^{14}$C background
is to avoid organic liquids and instead, to use   
liquified noble gas as a scintillator. Projects with helium, 
neon and xenon are under investigation. Argon and krypton
are not suitable because of their long-lived 
radioactive isotopes.
Tab.~\ref{tab:noblegas} summarizes features relevant
for the design of pp-neutrino detection.
 
\begin{table}[htb]
\caption{Characteristics of liquid noble gases as $\nu$-target
and detector medium.}
\label{tab:noblegas}
\newcommand{\m}{\hphantom{$-$}}
\newcommand{\cc}[1]{\multicolumn{1}{c}{#1}}
\begin{tabular}{llll}
\hline
   & He   & Ne & Xe \\
\hline
Boiling Tmp.[K]: & 4.2 & 27 & 165 \\
Z:               & 2 & 10 & 54 \\
Density [g/cm$^3$]: & 0.125  & 1.20  & 3.06 \\
Emission wl [nm]: & 73 & 80 & 175 \\
Photons/MeV: & 22000 & 15000?  & 42000\\
\hline
\end{tabular}\\[2pt]
\end{table}

High density liquid xenon, as proposed by the {\bf XMASS}
collaboration \cite{xmass}, 
provides efficient self-shielding.
A geometry similar to that of BX, but smaller in size, 
could thus be realized. The scintillation photons can be detected with  
photomultiplier technology at liquid xenon temperatures. 
Backgrounds 
can arise from $^{85}$Kr $\beta$-decay and from
$2\nu-\beta\beta$ decay of $^{136}$Xe. 
The first isotope must be less than 
$4\times 10^{-15}$g krypton/g. If the $\tau_{1/2}$ of $^{136}$Xe
is $<8\times 10^{23}$~years, then isotope separation is needed.
The best limits have
$\tau_{1/2}>1.1\times 10^{22}$~years  \cite{damataup01}. 
Since theoretical calculations
estimate $\tau_{1/2}\approx 8 \times 10^{21}$~years, 
the need for isotope separation is likely. Fig.~7
shows the spectrum of recoil electrons along  with the 
$2\nu-\beta\beta$ decay of $^{136}$Xe.
A prototype detector containing 100 kg of liquid Xe is under
construction. Photon detection will be achieved with 
newly developed low background photomultiplier tubes consisting  
of a steel housing and a quartz window.
Milestones during this R\&D phase
will include the determination of the $2\nu-\beta\beta$ half-life
and the optical properties of liquid xenon, such as the 
scattering length of scintillation light.

\begin{figure}
\label{fig:xespec}
\includegraphics[width=22pc]{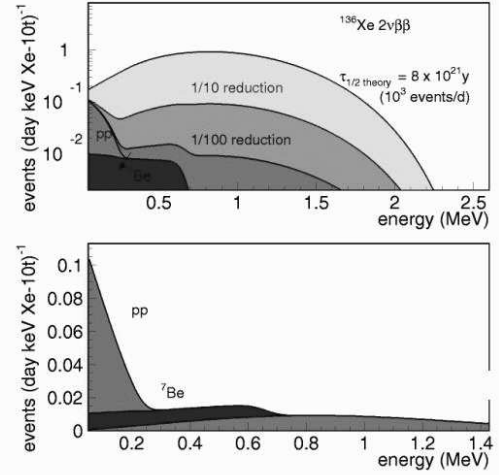}
\vspace{-1.5truecm}
\caption{Neutrino recoil spectrum displayed together with the 
$2\nu-\beta\beta$--decay of $^{136}$Xe. Isotope separation 
is required if  $\tau_{1/2}\approx 8 \times 10^{21}$~years.}
\end{figure}

A similar concept is being pursued by the {\bf CLEAN} project \cite{clean}.
The target under discussion is neon (or helium) for the fiducial
mass and liquid or solid neon as passive shielding buffer. 
Since neon scintillation occurs at 80 nm, a wavelength shifter 
is required in order to use photo multipliers for light detection.


\begin{figure}
\label{fig:heron}
\includegraphics[width=20pc]{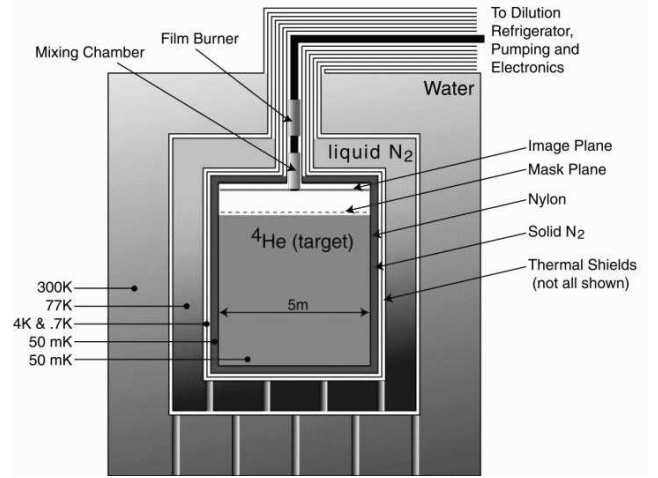}
\vspace{-1.0truecm}
\caption{Schematic view of HERON detector.}
\end{figure}

About 20t of super-fluid helium as neutrino target is modelled in the 
{\bf HERON} project \cite{heron}. 
Recoil electrons are detected via 1) prompt scintillation 
photons (80 nm) and 2) delayed phonon/rotons (1 meV; 10$^8$/MeV) 
on sapphire or silicon wafer calorimeters.
About 2400 of these wafers are proposed to be 
located above the liquid phase. A schematic view
of the detector is displayed in Fig.~8. 
The temperature rise after photon/phonon interactions
is sensed by super-conducting transition edge thermometers or  
by magnetic micro-calorimeters.  
The goal is to obtain single (16 eV) photon sensitivity per 
wafer. 
Internal radioactive backgrounds are not expected to be 
a problem, since super-fluid helium is ``self-cleaning'' 
(i.e. gravity~$>$~kT or chemical binding energies)
and impurities, if present, fall to the bottom. 
On the other hand, the low density of helium does not provide 
sufficient self shielding against external backgrounds. Therefore,
high radiopurity shielding material, dewar and light sensors 
is imperative. 
Point-like events, such as $\nu e^-$--scattering, can possibly 
be distinguished from multiple scattering events typical for external
$\gamma$--radiation by position reconstruction.

Full energy reconstruction of the neutrino energy in ES experiments
requires the determination the angle that the  recoil electron makes 
with respect to the incident neutrino direction. For this purpose {\bf
  TPC's} for solar neutrino detection are being investigated 
by several groups. The pioneering research project HELLAZ 
was one such project.
A follow up R\&D project was initiated recently \cite{tpc} that
proposes substantial conceptual modifications: it is planned to use methane
instead of isobutane to overcome the $^{14}$C background, to decrease the 
operational pressure so as to increase the angular resolution, and to simplify
the complexity of the electrodes in order to reduce the radioactive 
background.
A further project under discussion is {\bf Super-MUNU} \cite{munu}.
The detector concept is based on the experience of the MUNU experiment
that searched for a neutrino magnetic moment at the Bugey reactor. 
Four 50~m$^3$ modules, filled with 
CF$_4$ at 1-2 bars pressure are under discussion. 
As a prototype study, it has been proposed to 
operate the MUNU reactor detector 
in an underground laboratory in order to study radioactive 
background and detector performance (angular and energy resolution) 
at pp-$\nu$ energies.

\subsection{Charged current experiments for sub-MeV $\nue$ real
time detection}

The electron flavor content of sub-MeV solar neutrinos will be probed 
via neutrino capture. Only a few promising candidate nuclei exist 
for real time detection. 
Either the transition must populate an isomeric excited
state, or the final state needs to be unstable in order to provide
a coincidence tag to discriminate against background events.
Moreover, the energy threshold of the transition 
needs to be sufficiently low to have sensitivity to sub-MeV neutrinos.
Tab.~\ref{tab:realtimenuclei} lists nuclei which satisfy these condition.

\begin{table*}[htb]
\caption{Candidate nuclei for real time solar neutrino detection via
  neutrino capture (CC).}
\label{tab:realtimenuclei} 
\newcommand{\m}{\hphantom{$-$}}
\newcommand{\cc}[1]{\multicolumn{1}{c}{#1}}
\begin{tabular}{lllllll}
\hline
Isotope& Abundance & T$_{1/2}$  & Daughter & Q [keV]  & E$^{\ast}$ [keV] 
                     & T$^{\ast} _{1/2}$ \\
\hline
$^{115}$In & 95.7\% & $5\cdot 10^{14}$a  & $^{115}$Sn & 128 & 116+498 &3.26~$\mu$s \\
$^{176}$Yb & 12.7\% & stable  & $^{176}$Lu & 301 & 72 & 50 ns \\
$^{160}$Gd & 21.9\% & stable  & $^{160}$Tb & 244 & 75+64 & 6+60 ns \\
$^{82}$Se & 9.4\% & $1\cdot 10^{20}$a  & $^{82}$Br & 173 & 29 & 10 ns\\
$^{100}$Mo & 9.6\% & stable  & $^{100}$Tc & 168 & 3209 ($\beta$-decay
of ground state) &
15.8 s\\
$^{71}$Ga & 39.9\% & stable  & $^{71}$Ge & 404 & 175 & 79 ns\\
$^{137}$Ba & 11.2\% & stable & $^{137}$La & 611 & 11 & 89 ns\\
$^{123}$Sb & 42.7\% & stable & $^{123}$Te & 541 & 330+159 & 31 ns\\
$^{159}$Tb & 100\%  & stable  & $^{159}$Dy & 543 & 177+121+56 & 9.3 ns\\
\hline
\end{tabular}\\[2pt]
E$^{\ast}$ and T$^{\ast} _{1/2}$ refer to de-excitation energy and
half life of the isomeric state populate in the transition. 
\end{table*}

The {\bf MOON} collaboration \cite{moon} favors the molybdenum isotope 
$^{100}$Mo for solar neutrino detection. The threshold for neutrino 
capture is 168 keV and thus sensitive to pp-neutrinos. The GT
strength to the 1$^+$ ground state in $^{100}$Tc is measured to 
be $(g_A/g_V)^2 B(GT) = 0.52\pm 0.02 $ by EC and charge exchange reactions.
An interaction rate of 1.1 and 3.3 events per day and 10 tons of
$^{100}$Mo is expected for pp- and $^7$Be neutrinos. 
Fig.~9 displays the nuclear levels and transition
involved in the neutrino capture process. The neutrino signature
consists of a prompt electron and a delayed $\beta$-decay
($\tau=15.8$~s) with endpoint energy of 3.4~MeV.
Two different design options for a
detector based on 100 tons of natural 
molybdenum  (9.6 tons of $^{100}$Mo) exist:
1) a hybrid detector consisting of plastic scintillator bars
interleaved by Mo-foils (0.05 g/cm$^2$) 
and read out with PMTs via wave-length shifting fibers, 
or 2) a similar detector geometry  with liquid scintillators instead
of plastic. Fig.~10 shows the 
expected spectrum together with the background of $2\nu-\beta\beta$ decay. 

\begin{figure}
\label{fig:mo100_1}
\includegraphics[width=15pc]{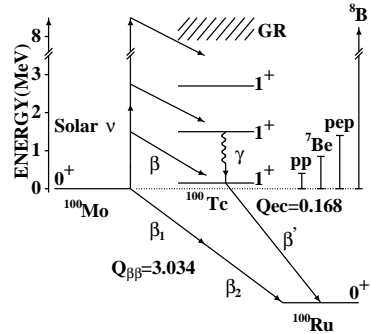}
\vspace{-2.truecm}
\caption{Level and transition scheme of $^{100}$Mo for solar neutrino
induced $\beta\beta^\prime$ decay \cite{moon}.}
\end{figure}

\begin{figure}
\label{fig:mo100_2}
\includegraphics[width=15pc]{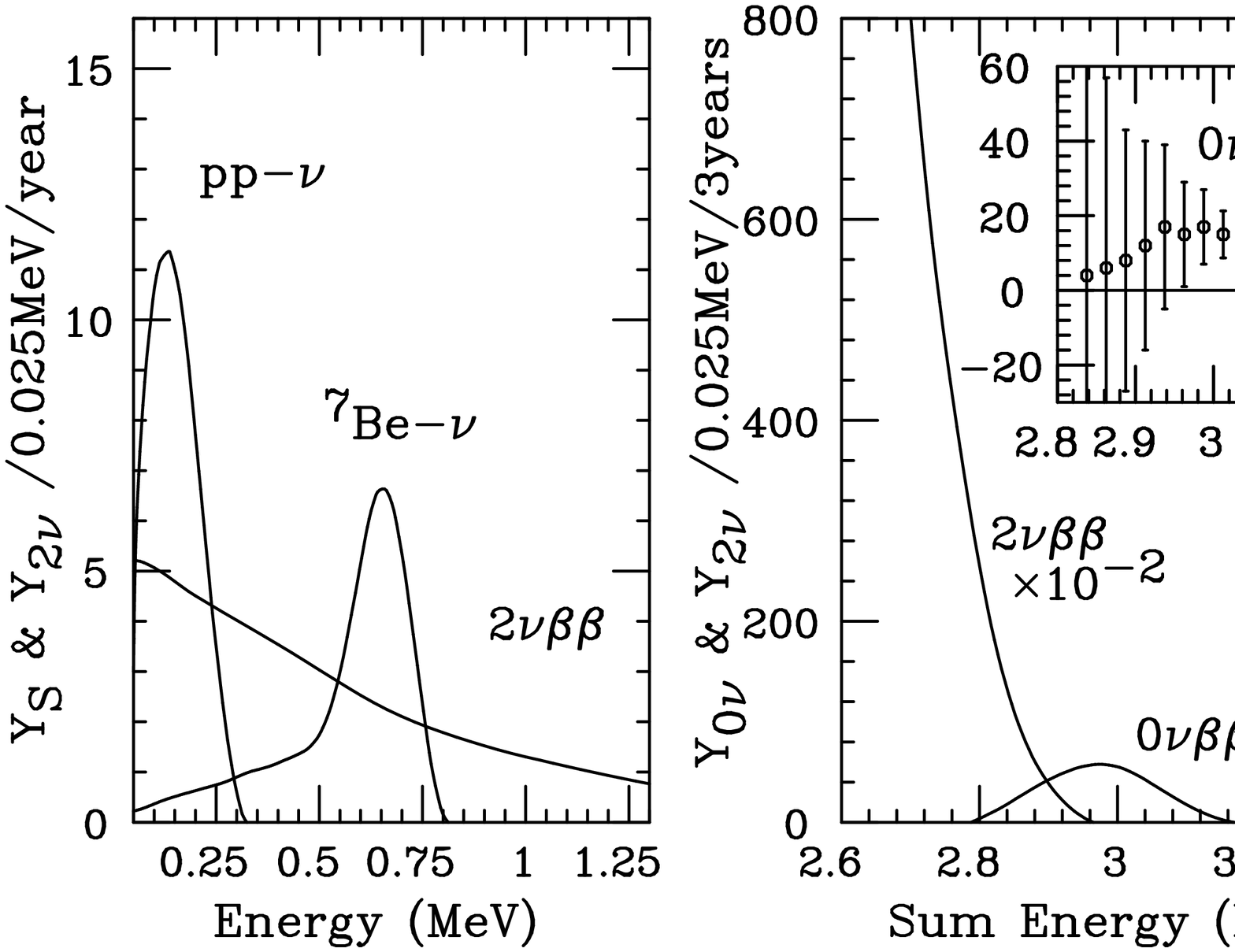}
\vspace{-1.truecm}
\caption{Energy spectrum for a possible detector with 3.3 tons of
$^{100}$Mo. The continuous line shows the expected background from 
$2-\nu\beta\beta$ decay  \cite{moon}.}
\end{figure}

Various candidate nuclei listed in Tab.~\ref{tab:realtimenuclei}
are being investigated by the {\bf LENS} collaboration \cite{lens}. 
These metals are being loaded up to 10\% in weight in 
organic liquid scintillators. Ligands under study are
carboxylic acids, phosphor organic compounds as well as
beta-diketonates. Among the various nuclei, $^{115}$In and $^{176}$Yb are
currently favored. Figs.~11 and 12  display 
the levels and transitions involved.   
$^{82}$Se and $^{160}$Gd are less promising because
of  challenges related to detection technique as well as 
to radioactive background. 

The $B(GT)$ transition strengths of $^{176}$Yb and  $^{115}$In 
have been measured via charge exchange reactions
in order to estimate the interaction cross section.
For $^{176}$Yb a $B(GT)$ value of $0.20\pm0.04$  ($0.11\pm0.02$) 
was evaluated for the transition to the 194.5 keV (339 keV) 
level \cite{BGTYb}. For $^{115}$In the weak matrix element 
is  $B(GT)=0.17$ \cite{BGTIn}. These values are sufficiently accurate
to evaluate the target mass necessary for the LENS detector.
However, to derive the neutrino fluxes, it is planned
to use an artificial $^{51}$Cr neutrino source  of several MCi
strength to calibrate the cross section at the few percent level.
Figs.~13 and 14 display the expected
spectra and interaction rates for 
20~t of natural ytterbium and 4~t of indium.
The actual target mass for an indium detector will most likely
be on the 10~t scale taking into account detector efficiencies
and required statistical accuracy.

\begin{figure}
\label{fig:Yblevels}
\includegraphics[width=15pc]{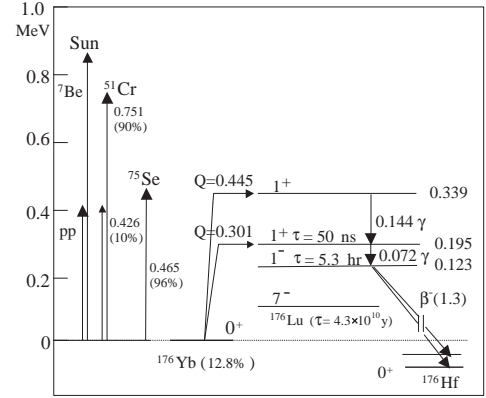}
\vspace{-1.truecm}
\caption{Level scheme and $\gamma$-ray tag for solar neutrino
detection by $^{176}$Yb. All energies in MeV \cite{BGTYb}.}
\end{figure}

\begin{figure}
\label{fig:Inlevels}
\includegraphics[width=12pc]{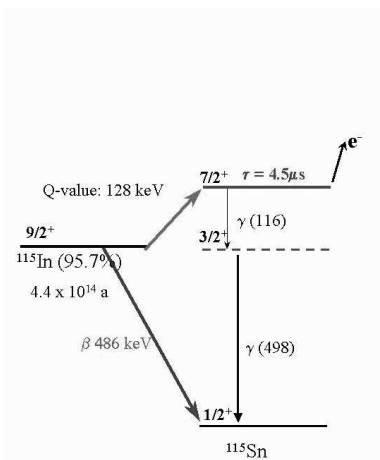}
\vspace{-1.5truecm}
\caption{Level scheme and $\gamma$-ray tag for solar neutrino
detection by $^{115}$In.} 
\end{figure}

A high granularity is required in order to discriminate 
against background caused by random coincidences of $^{14}$C 
and other sources. This will be realized by a modular detector
design. A basic module has a lengths of the order of the 
absorption length of the liquid scintillator which corresponds to 
a few meters. The modul area varies from $5\times 5$~cm$^2$ to
$20\times 20$~cm$^2$ depending on metal loading, scintillator 
performance, choice of  nucleus and neutrino source calibration
optimization.

In order to study the detector performance as close as possible to 
the final detector geometry, the LENS Low-Background-Facility (LLBF)
has been newly installed underground at the LNGS. A low-background passive 
shielding system with 80 tons of mass, located in a clean room, 
can house detector modules with dimensions up to 70~cm $\times$ 70~cm 
$\times$ 400~cm. All shielding materials have been selected in order 
to minimize the intrinsic radioactive contamination. First results 
from the prototype phase are expected in 2002. Of particular 
interest will be the experimental study whether pp-$\nue$ 
detection with $^{115}$In, as proposed in Ref.~\cite{raju-In} 
is feasible.

\begin{figure}
\label{fig:Ybspec}
\includegraphics[width=15pc]{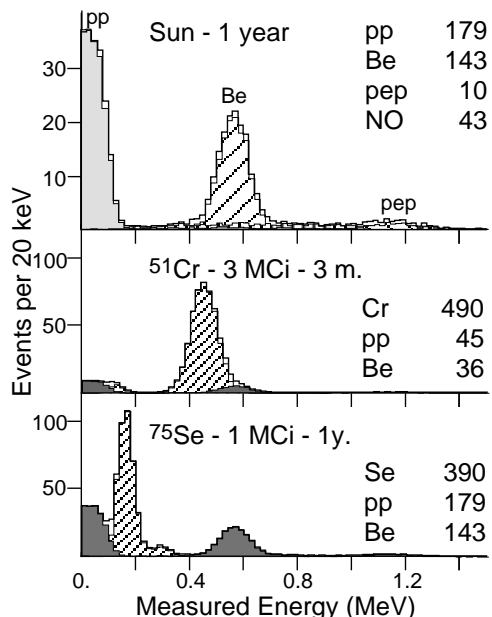}
\vspace{-1.0truecm}
\caption{Estimated $\nue$ spectra of an Yb loaded liquid scintillator
detector containing 20 tons of Yb (nat. abund.) exposed to 
a) solar neutrinos, b) radioactive $^{51}$Cr  and c)  $^{75}$Se 
$\nue$-source \cite{BGTYb}} 
\end{figure}

\begin{figure}
\label{fig:Inspec}
\includegraphics[width=15pc]{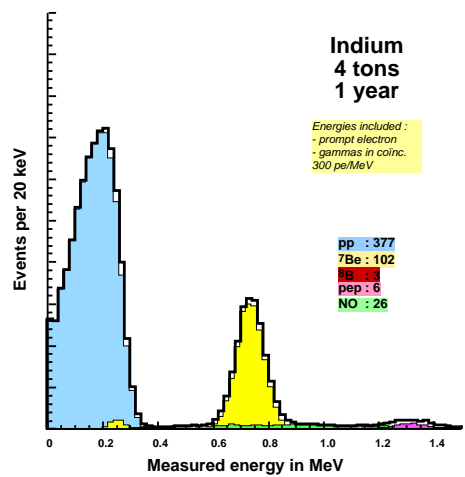}
\vspace{-1.0truecm}
\caption{Estimated $\nue$ spectra of an In loaded liquid scintillator
detector containing 4 tons of indium. The pp- and  $^7$Be rates are
377/year and 102/year.}
\end{figure}

\section{Study of $U_{e3}$ with reactor-$\nuebar$'s and other projects}

The goal of the  {\bf Kr2Det} project \cite{kr2det} is the study of 
$U_{e3}$ in a ``disappearance'' experiment using reactor-$\nuebar$'s.
Two identical detectors with 50 ton mass each, similar
 to the design of BOREXINO, are located at 150m and 1100m distant
from the reactor core. Both detectors will be located underground
at a depth of 600 mwe. From the comparison of the event rate
of both detectors,
a sensitivity of $|U_{e3}|^2 \le 3\cdot 10^{-3}$ (90\% CL) at
$\Dm2 = 3\cdot 10^{-3}$~eV$^2$ is expected. A prototype experiment
(MiniKr2Det) with 1.5 ton target mass is under design in order
to probe mixing parameters proposed by the LSND experiment.

The {\bf TEXONO} collaboration  in Taiwan \cite{texono} studies 
low-energy neutrino properties and interactions with 
reactor-$\nuebar$'s with  HP-Ge detectors and  CsI(Tl) scintillation 
cristalls crystals to search for a neutrino magnetic moment and
neutrino decay.

\section{Summary and outlook}

Solutions with $\Dm2 < 10^{-6}$ eV$^2$ {\em or} small mixing 
angles can be probed best with solar neutrino experiments at 
sub-MeV energies. Values of $\Dm2 > 10^{-6}$ eV$^2$ 
{\em and} large mixing angles can be studied both with long-baseline nuclear 
reactor neutrino experiments  and with sub-MeV 
solar neutrinos. Large mixing, in particular at large $\Dm2$ values,
is favored when combining all solar neutrino data. 
In particular, the LMA solution with $\Dm2 = 3.7  \times
10^{-5}$~eV$^2$ 
and $\tan^2 \Theta = 0.37$ \cite{concha} gives the 
best $\chi^2$ values in global analysis. 
The exact value of the minimum, as well as the 
extension of the allowed parameter range, depend mainly 
on the treatment of the $^8$B flux in the analysis. 
A large part of the $10^{-4}$~eV$^2$ range is allowed by the current data. 
Ultimately,  large $\Dm2$ values are limited  at $10^{-3}$~eV$^2$
by the CHOOZ reactor experiment \cite{chooz}.

Results of the upcoming experiments KamLAND and BOREXINO
most likely will tell us which oscillation solution is realized
in nature. 
The objective of the upcoming experiments will be to probe the 
mixing parameters with complementary
techniques both with ES and CC reactions {\em with high precision}. 
For the LMA solution, reactor and solar neutrino experiments 
are complementary to determine mixing parameters:
$\Dm2$ can be derived with high precision from reactor 
neutrino experiments while the mixing angle is measured 
best with solar neutrinos.  A direct measurements of the 
primary solar neutrino fluxes will be essential 
to probe, with high accuracy, solar model predictions. 
Perhaps even new surprises await, if so, it would not be
the first time in astroparticle physics -- new experiments
may find new phenomena.

\vspace{1.truecm}
\noindent {\bf Acknowledgement}

It is a pleasure to thank all colleagues and friends
with whom I had fruitful discussions while preparing this short 
review of this very active and exciting field, 
and who provided me with the latest
developments of their projects. In particular, 
I would like to thank F.X. Hartmann for the careful 
reading of the manuscript.


\begin{thebibliography}{9}

\bibitem{till} T. Kirsten, Rev. Mod. Phys., 71(1999) 1213-1232.

\bibitem{concha}  recent golabal analysis can be found in: 
  J.N. Bahcall, M.C. Gonzalez-Garcia, Carlos
  Pena-Garay, hep-ph/0111150v2; P.I. Krastev and A.Yu. Smirnov, 
  hep-ph/0108177.

\bibitem{kamland} http://www.awa.tohoku.ac.jp/KamLAND/;
                  Proposal for US Participation in KamLAND (1999) 
                  http://kamland.lbl.gov/ KamLAND.US.Proposal.pdf.

\bibitem{borexino} BOREXINO Collaboration, G. Alimonti et al., 
Astrop. Phys. 16, (2002) 205-234. Also at hep-ex/0012030. 

\bibitem{stefanTaup97} S. Sch\"onert, TAUP 1997, Nucl. Phys. B (Proc. Suppl.)
         70 (1999) 195-198.


\bibitem{lens} R.S. Raghavan, PRL Vol.78 No.19 (1997) 3618;
               LENS collaboration, Letter of Intent, Laboratori Nazionali del 
               Gran Sasso (1999);
               


\bibitem{xmass} Y. Suzuki, LowNu2 workshop, Tokyo, 2000, 
                http://www-sk.icrr.u-tokyo.ac.jp/neutlowe/,
                S.Moriyama, M.Yamashita, XENON01, 2001,
                http://www-sk.icrr.u-tokyo.ac.jp/xenon01/index.html


\bibitem{McDonald} A. McDonald, see this proceedings;
                 Q.R. Ahmad et al. (SNO collaboration) PRL 87 (2001)
                 071301.

\bibitem{chooz} M. Apollonio et al., Phys. Lett. B 466 (1999) 415.

\bibitem{BahKrastSmir}J.N. Bahcall, P.I. Krastev, A.Y. Smirnov, JHEP
  0105 (2001) 015


\bibitem{clean} D.N. McKinsey and J.M. Doyle, J. of Low Temp. Phys.,
  118, 153 (2000).

\bibitem{heron} B. Lanou, LowNu2 workshop, Tokyo, 2000, 
                http://www-sk.icrr.u-tokyo.ac.jp/neutlowe/,
http://www.sns.ias.edu/~jnb/Meetings/ Lownu/lownuPres/pindex.html.

\bibitem{moon} H. Ejiri et al., PRL 85 (2000) 2917-2910

\bibitem{tpc} G. Bonvicini et al., Contributed paper, Snowmass 2001,
                 hep-ex/0109199, hep-ex/0109032.

\bibitem{munu} Super-MUNU, http://isnwww.in2p3.fr/ munu/Supermunu/solar.html;
               C. Broggini et al., LowNu workshop,  Sudburry (2000).
 

\bibitem{genius} L. Baudis, H.V. Klapdor-Kleingrothaus, Proc. Beyond the 
        Desert'99 (1999), astro-ph/0003435; 
        H.V. Klapdor-Kleingrothaus, L. Baudis, G. Heusser, 
        B. Majorovits, H. Paes, hep-ph/9910205.

\bibitem{lowc14} E. Resconi, Dissertation Univ. di Genova (2000);
                 S. Sch\"onert and E. Resconi, to be published;
                 R.S. Raghavan, unpublished.

\bibitem{kr2det} Y. Kozlov, L. Mikaelyan and V. Sinev, 
                NANP-2001, Dubna (2001), hep-ph/0109277
        
\bibitem{texono} C.Y. Chang et al., Nucl. Phys. B (Procs. Suppl.)
                 66, 419 (1998), H.T-K. Wong and J. Li, NCTS workshop 2001, 
                 hep-ex/0201001.

\bibitem{Em4} S. Sch\"onert, T. Lasserre, L. Oberauer, to be
              submitted; S. Schoenert, NOON01, 
              http://www-sk.icrr.u-tokyo.ac.jp/noon2001/

\bibitem{BXC14} G. Alimonti et al., BOREXINO collaboration,
                Phys. Lett. B 422 (1998) 349. 


\bibitem{damataup01} DAMA collaboration, TAUP01, this proceedings.


\bibitem{BGTYb} M. Fujiwara et al., PRL 85, 4442 (2000);
                  M. Bhattacharya et al., PRL 85, 4446 (2000).

\bibitem{BGTIn} J. Rappaport et al. PRL 54, 2325 (1984).

\bibitem{raju-In} R.S. Raghavan, hep-ex/0106054 (2001).

\end{thebibliography}
\end{document}